\documentclass[11pt]{article}
\usepackage[total={5.5in, 9in}]{geometry}
\usepackage{url}
\usepackage{doi}

\usepackage{hyperref}
\usepackage[nameinlink, capitalise]{cleveref}
\usepackage{graphicx}
\begin{document}
\title{Estimating the Peer Degree of Reachable Peers in the Bitcoin P2P Network}
%
%
\author{Matthias Grundmann, Max Baumstark, Hannes Hartenstein}
%
%
\date{Institute of Information Security and Dependability (KASTEL) \\
Karlsruhe Institute of Technology (KIT), Karlsruhe, Germany }
%
\maketitle              
\begin{abstract}
A recent spam wave of IP addresses in the Bitcoin P2P network allowed us to estimate the degree distribution of reachable peers in the network.
The resulting distribution shows that about every second reachable peer runs with Bitcoin Core's default setting of a maximum of 125 concurrent connections and nearly all connection slots are taken.
We validate this result and, in addition,  use our observations of the spam wave to group addresses that belong to the same peer.
By doing this grouping, we improve on previous measurements and show that simply counting addresses overestimates the number of reachable peers by 13\,\%.
\end{abstract}

\section{Introduction}

To join the Bitcoin P2P network, new peers need to learn the addresses of peers that are already part of the network.
A peer has one or multiple addresses that can be used to find the peer and initiate connections to the peer.
Bitcoin uses a decentralized way to disseminate addresses to peers:
A peer announces its own addresses by sending \textsc{addr} messages to its neighbors and the neighbors forward the addresses to other peers.
In July and August 2021, a huge wave of addresses was flooded in the Bitcoin P2P network that caused an increase in the number of addresses distributed per day from 40,000 to about 6,000,000 unique addresses per day \cite{web-dsn-bitcoin}.
These spam addresses did not belong to actual peers and were sent by an unknown party.
While we do not know the purpose for sending the spam addresses, we look at the effects that the spamming had and what information about the topology of the Bitcoin P2P network can be extracted from observing the effects.
We estimate the degree (number of neighbors) of reachable peers.
While previous work has shown that the peer degree distribution of other cryptocurrencies' P2P networks resembles a power law distribution \cite{delgado-segura_txprobe_2019,cao_exploring_2020,wang_ethna_2021-1}, our observations indicate that in the Bitcoin P2P network about half of the peers have a degree of around 125.
Because 125 is the default maximum for connections of Bitcoin Core, the most commonly used client, this finding means that many peers do not have slots available for new incoming connections.
As the ability for peers to connect to other peers in the network is important for the health and resilience of the P2P network, we validate this observation by running an experiment to measure how many peers accept incoming connections and find that more than 50\,\% of all reachable peers do not accept additional incoming connections or are close to their connection limit.
We further show that the majority of peers being hosted in the networks of cloud providers have around 125 connections while the networks of ISPs include peers that tend to have fewer neighbors.
Finally, we estimate the number of unreachable peers in the Bitcoin P2P network from the peer degree distribution.
We estimate that there are about 32,800 unreachable peers in the network which aligns with estimations from previous work \cite{neudecker_security_2019,web-lukejr-history,grundmann_announcements_2022}.
and find sets of addresses that belong to the same reachable peers.
This mapping shows that estimating the number of reachable peers by counting reachable addresses overestimates their number by about 13\,\%.

{\em Related Work.}
While different methods to learn about the topology of the Bitcoin P2P network have been proposed, most of them were impractical or too costly to be run in the real Bitcoin P2P network.
A notable exception is AddressProbe \cite{miller_discovering_2015} that exploited an information leak in the handling of addresses to infer connections between reachable peers.
The authors of \cite{miller_discovering_2015} used AddressProbe to infer the topology of P2P network's subgraph that contains only reachable peers and calculate the resulting peer degree distribution.
The degree distribution showed that the majority of reachable peers had a degree between eight and twelve which differs strongly from our results because our results also include connections between reachable and unreachable peers.
While other methods to infer parts of the topology have been proposed \cite{neudecker_timing_2016,grundmann_exploiting_2019,delgado-segura_txprobe_2019}, these methods were too costly to be run in the Bitcoin P2P network.
However, the peer degree distribution of other cryptocurrencies' P2P networks has been analyzed, e.g. the P2P networks of the Bitcoin testnet \cite{delgado-segura_txprobe_2019}, Monero \cite{cao_exploring_2020}, and Ethereum \cite{wang_ethna_2021}.
The Bitcoin transaction network, sometimes simply referred to as the `Bitcoin network', is the graph defined by the transactions of the Bitcoin blockchain.
The topology of this network has been analyzed previously \cite{ron_quantitative_2013,reid_analysis_2013,lischke_analyzing_2016,filtz_evolution_2017,di_francesco_maesa_data-driven_2018,tao_complex_2021-1} but the transaction network is completely different from the Bitcoin P2P network which is the focus of this work.

\section{Observations and Monitoring Setup}

In July 2021, user piotr\_n reported in the BitcoinTalk Forum \cite{web-bitcointalk} that spam addresses were distributed in the Bitcoin P2P network.
piotr\_n found that the behavior of the spamming peers is to connect to reachable peers, send them 500 \textsc{addr} messages with ten spam addresses each, and then disconnect.
We observed the behavior described by piotr\_n at a reachable peer:
During July and August 2021, about 400 times one of this peer's neighbors sent within a few seconds a batch of 5,000 unique IPv4 addresses.
Over the observed time, the spam originated from 243 different IP addresses.
All spam addresses in a batch had the same associated timestamp which was set to a value up to nine minutes into the future.
We analyzed the distribution of the received spam addresses and found that they were distributed uniformly over the IPv4 address space and included IP addresses from reserved IPv4 address blocks like 127.0.0.0/8.
We take this finding as evidence that the spam addresses were randomly chosen and did not belong to actual peers.

Our monitoring setup consists of three monitor nodes that connect to all reachable peers but do not accept incoming connections.
Two of those monitor nodes are located in the network of our university (AS 34878) and a third monitor node is located in a different location (AS 680).
All monitor peers log received \textsc{addr} messages and connections to other peers that are opened or closed.

\section{Estimating the Degree of Reachable Peers}

\begin{figure}[tbp]
  \centering
  \includegraphics[width=\linewidth]{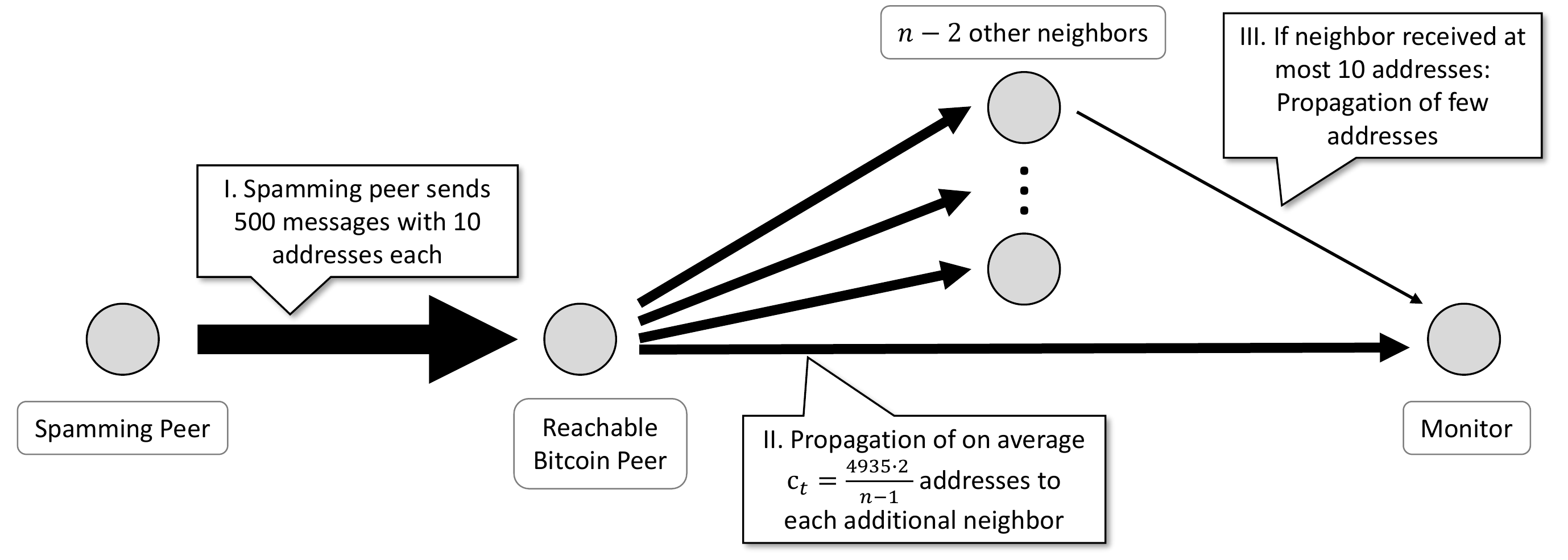}
  \caption{Overview of the peer degree estimation. A reachable peer is connected to a spamming peer, our monitor and $n-2$ other peers. The spamming peer sends $500 \cdot 10$ addresses to the reachable peer (I). The peer propagates the addresses to all neighbors except the spamming peer (II). From the number of propagated addresses, the monitor can estimate the number of neighbors.
       }
  \label{fig-degree-estimation}
\end{figure}

Bitcoin Core, the Bitcoin reference client that is used by most peers \cite{web-bitnodes}, accepts addresses with an associated timestamp of up to ten minutes into the future and propagates addresses until their associated timestamp is older than ten minutes.
Additionally, an address is only propagated if it was received in an \textsc{addr} message with a size of at most ten.
Because both conditions are fulfilled for the spam addresses, a peer that runs Bitcoin Core considers these addresses for propagation to its neighbors.
However, Bitcoin Core forwards only routable addresses and about 1.3\,\% of the IPv4 address space are considered as unroutable \cite{web-github-is-routable}.
Therefore, on average $4,935$ addresses of the $5000$ received addresses are forwarded.
Because each routable address is forwarded to two peers but not the peer that the address was received from, a peer $p$ with $n_p$ neighbors forwards each address to two out of $n_p-1$ neighbors and sends on average $c_p = 4,935 \cdot \frac{2}{n_p-1}$ addresses to each neighbor.
Consequently, our monitor nodes receive on average $c_p$ addresses from each peer that receives 5,000 spam addresses and we can estimate the number of neighbors of each reachable peer based on these observations (see \cref{fig-degree-estimation}).
While the main idea of this estimation approach has been proposed in 2014 by Biryukov et al. \cite[Section 10.1]{biryukov_deanonymisation_2014-1}, to the best of our knowledge results of this method applied to the Bitcoin P2P network have so far not been published.

\subsection{Estimation and Validation}

Our monitor nodes are connected to each reachable peer and receive the propagated spam addresses (see \cref{fig-degree-estimation}, II).
However, our monitor nodes also receive spam addresses that are not directly forwarded from a spamming peer (\cref{fig-degree-estimation}, III).
To filter out these messages and get only directly forwarded messages, we (1) analyze only \textsc{addr} messages received at the monitor that contain at least four entries and (2) we select only those addresses that have a timestamp that is three to ten minutes into the future from the point when the \textsc{addr} message was received and (3) we analyze only addresses if $c_{p,t}$, the number of addresses we received with the same timestamp $t$ from peer $p$, is greater than ten.
For each batch of spam address messages with size $c_{p,t}$, we calculate $n_{p,t} = 1 + 4,935 / c_{p,t} \cdot 2$ as an intermediate estimate for the number of neighbors of peer $p$.
As the intermediate estimates contain outliers, we calculate the estimate $n_p$ for the number of neighbors of peer $p$ by determining the median of all intermediate estimates $n_{p,t}$ during the time window of one day.
The length of this time window is chosen as a trade-off between a short time window during which the number of a peer's neighbors remains constant and a longer time window during which we collected more observations to get a more precise estimate.

To validate the estimation approach, we logged at three reachable validation peers the number of neighbors and compared the logs to our estimation.
Two of the three validation peers received the spam address tuples on their IPv4 and IPv6 address, the other peer only on its IPv4 address.
For each of these five addresses of our validation peers and each day with observed spam, we estimate the peer's degree using the above method and compare it to the peer's connection count logs.
As ground truth we take for each peer the average connection count of this peer during this day.
We compute the mean deviation of each estimate from this ground truth in percent and average the absolute percentage values.
This calculation leads to an average deviation of 4.1\,\% which means that the estimation is reasonably reliable.

\subsection{Resulting Degree Distribution}

\begin{figure}[tbp]
  \centering
  \includegraphics[width=\linewidth]{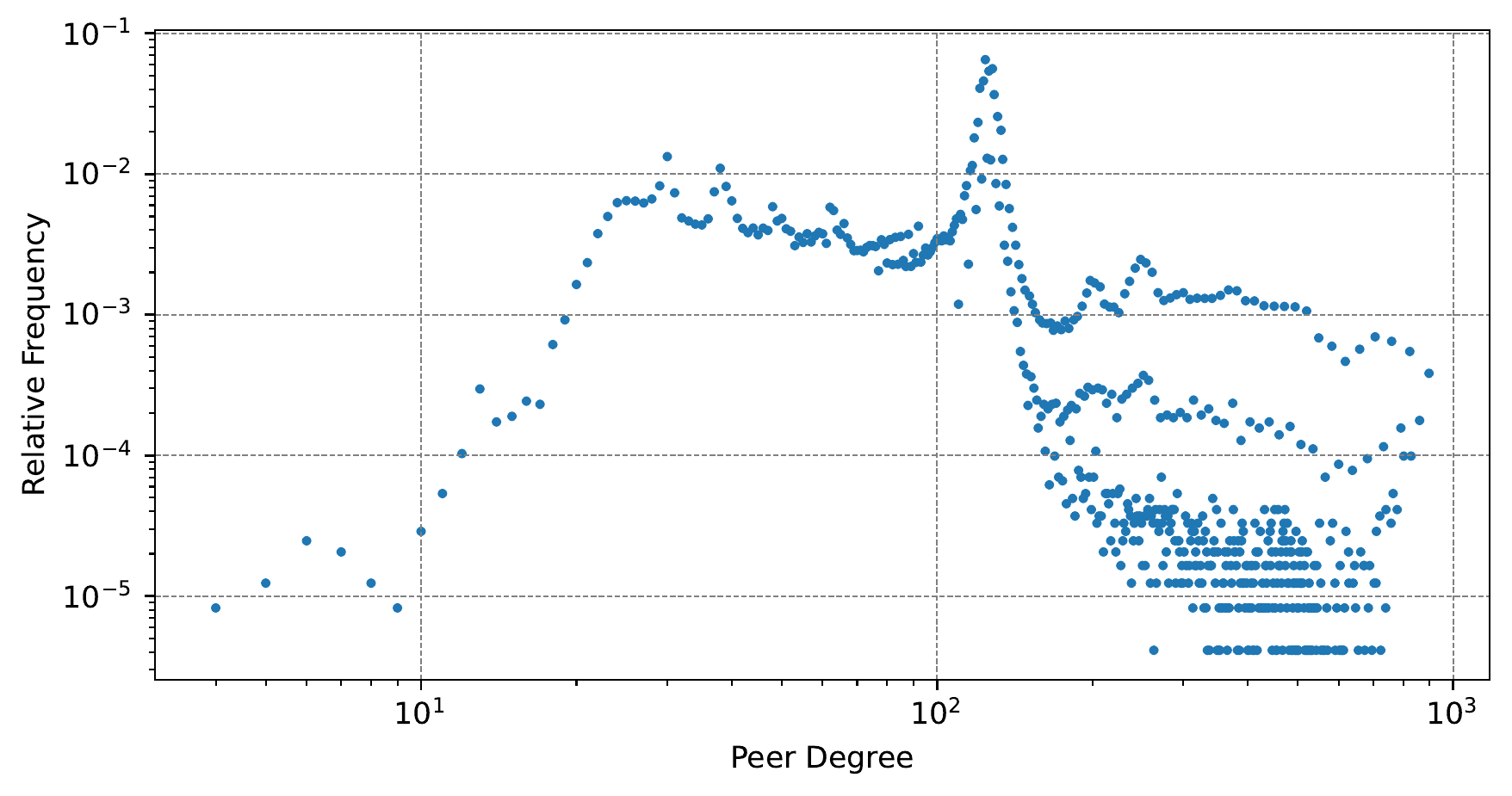}
  \caption{Relative frequencies of estimated peer degrees.}
  \label{fig-degree-scatter-all}
\end{figure}

\begin{figure}[tbp]
  \centering
  \includegraphics[width=\linewidth]{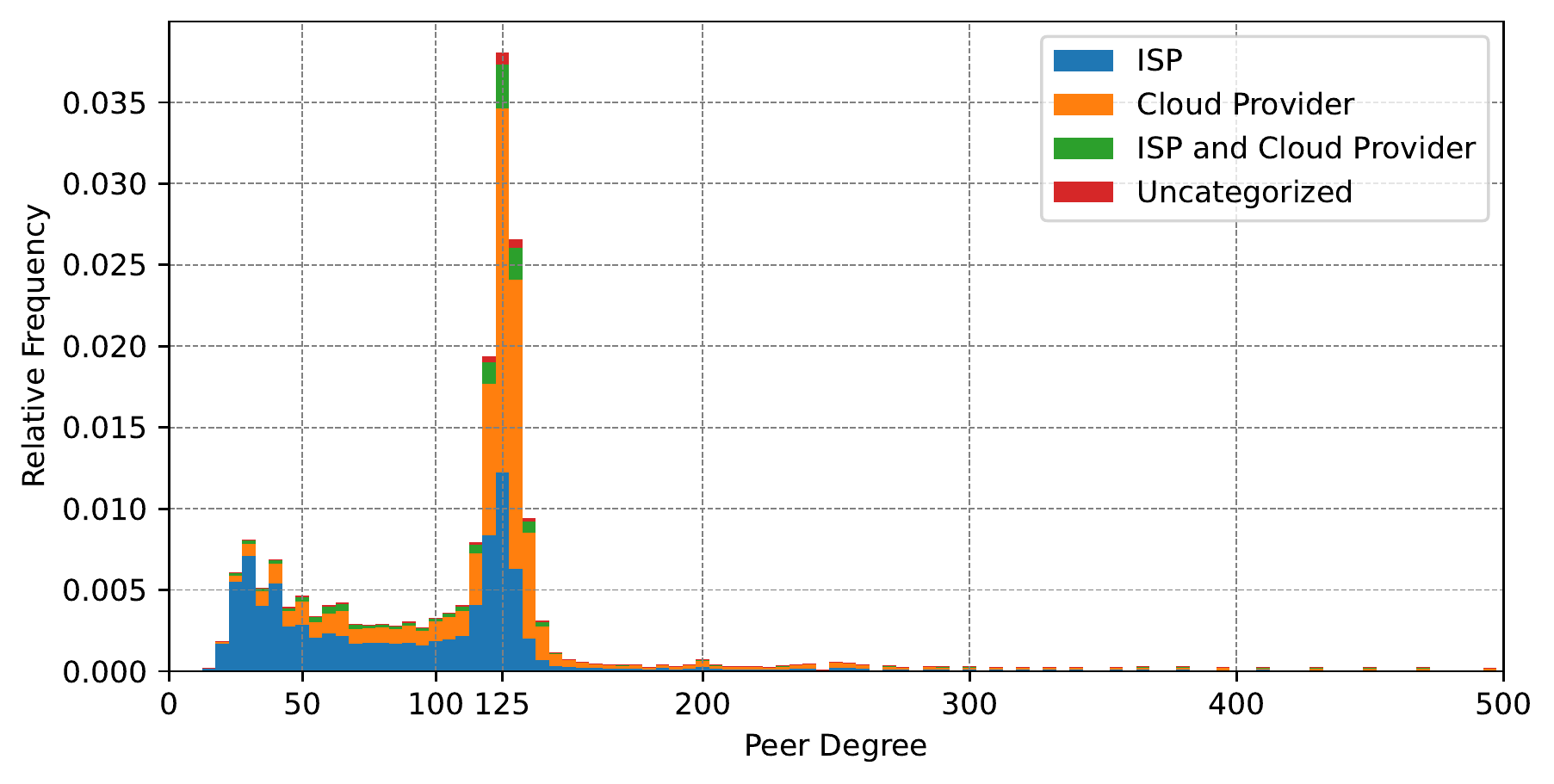}
  \caption{Normalized histogram (with a bin width of 5) of the estimated peer degree of all reachable peers.
    The colors indicate our categorization of the autonomous system that peers are located in.}
  \label{fig-degree-histogram-stacked}
\end{figure}

One important topological characteristic of a network is the distribution of peer degrees which can be estimated based on our observations.
Using the method described above, we receive one estimate per peer per day.
We show the resulting distribution of peer degrees in \cref{fig-degree-scatter-all}.
The distribution shows that the majority of reachable peers has an estimated degree of around 125, which is the default maximum number of connections in Bitcoin Core.
These results suggest that about 50\,\% of the reachable peers use this default configuration and all of their connection slots are filled.
The distribution of estimated peer degrees has a long tail of a few peers that have more than 140 connections.
We suppose that there are reachable peers with even more connections, however, we can only estimate the degree of peers with up to 1,000 connections.\footnote{This restriction is due to the fact that we use the condition that $c_t > 10$ to distinguish between addresses that were directly forwarded after being received from a spamming peer and addresses that were received from another peer. With the knowledge of which spam addresses are sent to which peer, this restriction is not necessary and higher degrees can be estimated, too. }

We also looked up the autonomous system (AS) of each peer's address using \cite{web-asn-mapping} and categorize each AS into the four categories `ISP', `Cloud Provider', `Both', and `Uncategorized'.
We manually classified ASes that contain a large percentage of peers and retrieved the category for the remaining ASes from the ASdb \cite{ziv_asdb_2021} database.
\Cref{fig-degree-histogram-stacked} shows the distribution of peer degrees separated by the category of a peer's AS.
While the median of estimated degrees for peers hosted at cloud providers is 125, the median of estimated degrees for peers located in networks by ISPs is 97.
This result shows that most peers with high degrees are hosted by cloud providers while the majority of peers with low degrees are located in the networks of ISPs.
One reason might be that peers running in data centers accumulate more incoming connections because they are less often restarted and their addresses are better distributed in the network because they change their address less often than peers running outside of data centers.

\subsection{Measurement of Available Slots for Incoming Connections}

Above results show that many reachable peers maintain the default maximum amount of connections and do not have slots for incoming connections available.
We validate this result by running the following experiment.

A reachable peer running Bitcoin Core always accepts a new incoming connection but, if the new connection fills the last remaining connection slot, a connection is evicted.
The evicted connection might be the connection that was just accepted but it might also be a previously existing connection.
To harden the resilience against Eclipse attacks, some connections are protected from eviction.
The remaining connections are grouped based on their AS and the youngest connection from the AS with the most connections is evicted.

In our experiment, we run a test peer that walks through a list of all reachable peers and opens a TCP connection to each peer.
If a connection was established, the test peer waits for three seconds and checks if the connection is still open.
If it is, the test peer opens four additional TCP connections to this peer, waits for three seconds and checks whether all five connections are still open.
Based on the behavior of Bitcoin Core described above, we expect the following results:
\begin{itemize}
  \item If a peer $p$ has more than five incoming connections slots available, the peer $p$ accepts all five tested incoming connections.
  \item If a peer $p$ has no incoming connection slots available and the test peer's AS is the AS with the most incoming connections to peer $p$, the peer $p$ evicts the first incoming connection.
  \item If a peer $p$ has no incoming connection slots available and there is an AS from which more peers are connected to $p$ than from the test peer's AS, the peer $p$ accepts the first new connection and evicts another connection. When the four additional connections are opened, the test peer's AS might become the AS with the most connections and the peer $p$ evicts our test peer's connections which means that we would see the first connection accepted but some of the additional connections evicted.
\end{itemize}

We run the experiment in November 2021 from three test peers located in two different AS.
To create the list of reachable peers, we collected all addresses that we received in unsolicited \textsc{addr} messages at one of our monitors on the day before.
Our test peers were able to connect to on average 9,461 peers of which 4,493 (47\,\%) accepted all five incoming connections.
On average 2,360 (25\,\%) accepted the first connection but not all five connections and 2,608 (28\,\%) evicted already the first connection.
We conclude that for 28\,\% of the reachable peers the slots for incoming connections are all taken while 25\,\% of the reachable peers are close to their capacity.
Only 47\,\% of the reachable peers seem to freely accept incoming connections.
This result confirms our interpretation of the peer degree distribution and shows that slots for incoming connections are a limited resource.

\section{Finding Peers with Multiple Addresses}

\begin{figure}[tbp]
  \centering
  \includegraphics[width=\linewidth]{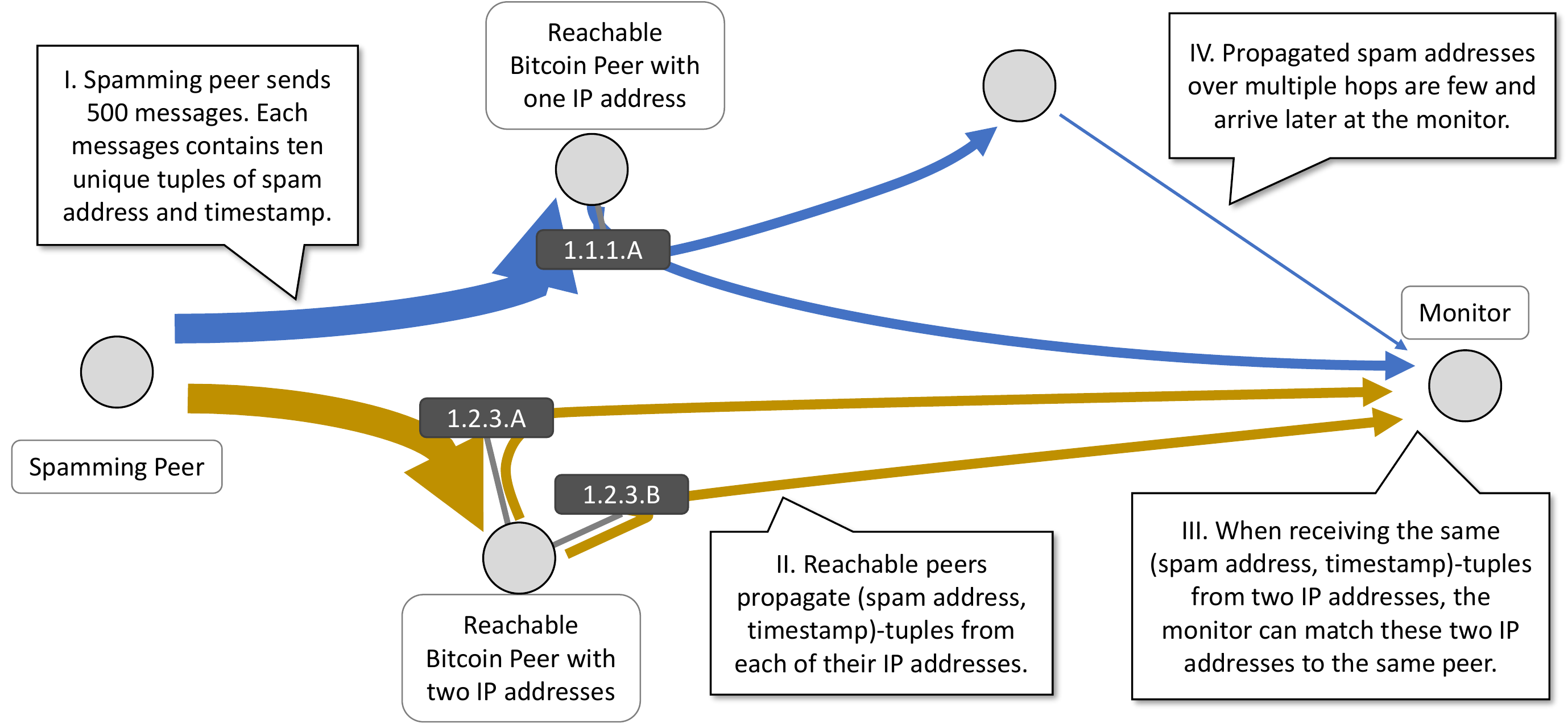}
  \caption{A peer with multiple reachable IP addresses is connected to the monitor with each IP address. By observing the propagated spam, the monitor can find IP addresses that belong to the same peer.
       }
  \label{fig-multiple-addresses}
\end{figure}

As far as we observed, the 5,000 spam addresses that are sent to one peer with one timestamp are not sent to another peer with the same timestamp.
Thus, a batch of spam addresses with the same timestamp mark a specific peer and we can use these markers to find multiple IP addresses that belong to the same peer (see \cref{fig-multiple-addresses}).
A reachable peer with multiple IP addresses that received spam addresses from a spamming peer forwards the spam addresses on all of its IP addresses (\cref{fig-multiple-addresses}, II).
If tuples of spam address and timestamp were received by the monitor from two different IP addresses, we can match these IP addresses to the same peer (III).
False positives can occur if spam addresses are propagated over multiple hops (IV).
To filter out these indirectly received spam addresses, we ignore spam addresses that have a timestamp less than five minutes into the future or if they were received from an IP address that sent us fewer than ten spam addresses with the same timestamp.
Further, we only match two IP addresses to the same peer if at least five identical tuples of spam address and timestamp were received by the monitor from both IP addresses.

We run this analysis on data collected by our monitor nodes and obtain sets of addresses that belong to the same peers.
After merging all intersecting sets of addresses that belong to the same peer, we get a mapping of 3,478 addresses to 1,536 peers.
While there seems to be one peer having 286 IPv6 addresses of the same /118 subnet, the majority of peers has only two addresses.
Most of these pairs of addresses are an IPv4 and IPv6 address, however, there are some pairs that are both IPv4 or IPv6 addresses. 
We validate the method using three of our peers that are using an IPv4 and IPv6 address and find that their IP addresses were correctly matched.
While one can estimate the number of reachable peers by counting reachable addresses, we can improve such estimations using the mapping from addresses to actual peers:
In August 2021, our monitor nodes were connected on average to 8,800 reachable IP addresses per day which map to 7,650 unique reachable peers per day.

To cross-check with our estimation of the peer degree, we compare the peer degrees for each set of matching addresses matched to the same peer and find that matching peers have very similar estimated degrees:
We estimate the average peer degree over the whole observed time span for each address.
For each set of addresses that we matched to the same peer $p$, we calculate the average degree $\tilde{n}_p$ of the estimates for the addresses of $p$.
Then, we calculate the relative deviation of the estimate for each address of $p$ from the mean $\tilde{n}_p$.
We determine this deviation between the estimates for each peer and find that the average over all deviations is only 0.2\,\%.

\section{Estimating the Number of (Unreachable) Peers}

\begin{figure}[tbp]
  \centering
  \includegraphics[width=\linewidth]{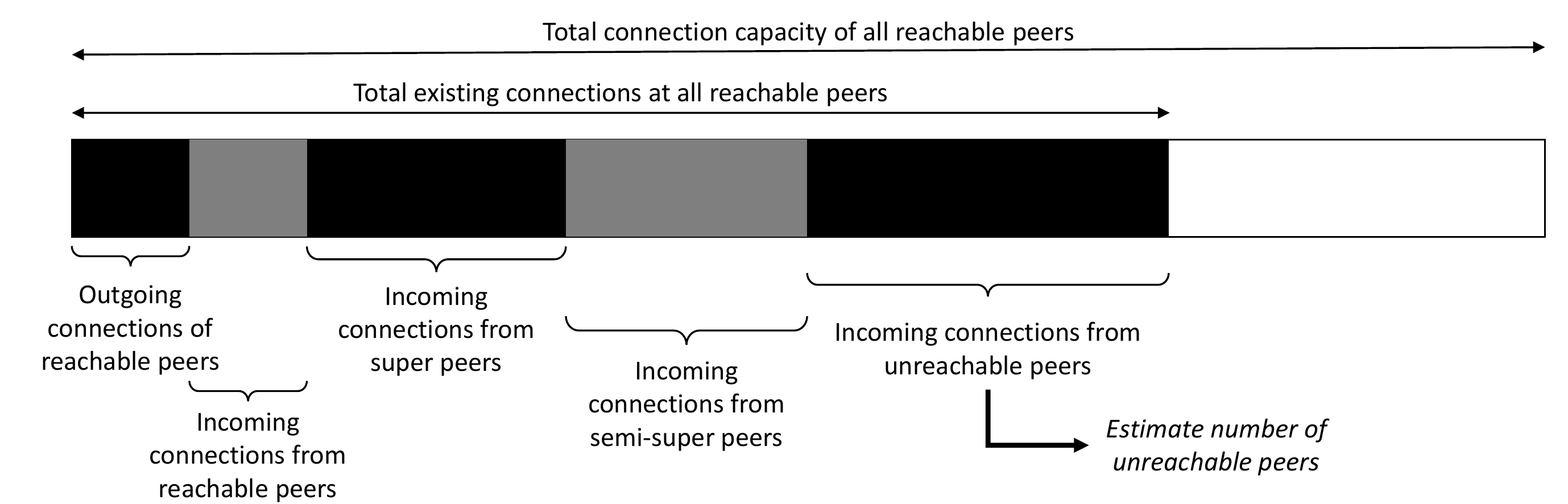}
  \caption{Usage of connection slots of all reachable peers.
       }
  \label{fig-unreachable-distribution}
\end{figure}

Whether they are reachable or unreachable, peers in the Bitcoin P2P network create outgoing connections to reachable peers.
If each peer would create exactly ten outgoing connections and we knew the number of incoming connections at reachable peers, we could infer the total number of peers by dividing the number of incoming connections at reachable peers by ten.
While we can estimate the number of existing (incoming and outgoing) connections at reachable peers from the peer degree distribution, not every peer in the real network creates exactly ten outgoing connections.
For instance, there are peers such as our monitor nodes that create outgoing connections to all reachable peers and, thus, have multiple thousand outgoing connections.
We call these peers super peers.
We further assume a class of semi-super peers that open connections to half of the reachable peers to model peers that open connections to many but not all of the reachable peers.

We have estimated above that the number of reachable peers is about 7,650 peers.
Assuming that each reachable peer runs Bitcoin Core with the default configuration, each reachable peer opens ten connections and, thus, we estimate that there are $7,650 \cdot 10$ outgoing connections of reachable peers.
As every outgoing connection is an incoming connection at another peer, there are also $7,650 \cdot 10$ incoming connections from reachable peers.

To find the number of super peers, we analyze the logs of three reachable peers that have been running for several months.
In October and November 2021, on average $18$ peers were connected to the three reachable peers.
Therefore, we assume that there are $18$ super peers that are connected to all reachable peers in the network.
To estimate the number of semi-super peers, we count the number of peers that were connected to two reachable peers that do not have a connection limit.
On average 44 peers were connected to both of these two reachable peers.
Therefore, we assume a number of $44 - 18 = 26$ semi-super peers.
The super peers take up $18 \cdot 7,650$ connection slots and the semi-super peers take up $26 \cdot 7,650/2$ connection slots in the network.

To estimate how many connections exist in the network, we calculate the sum of all estimated peer degrees of peers that have an estimated degree not higher than $130$.
We use this cut-off of the default maximum of $125$ and an error margin of 4\,\% and ignore peers with a higher degree because for peers with a higher degree we know that they are not using the default configuration and we do not know how many of their connections are outgoing or incoming connections.
Using our estimated peer degree distribution, we get an estimate of $712,840$ filled connection slots.
Subtracting the number of incoming connections that we ascribe to reachable peers, super peers, and semi-super peers, we get a remaining number of $322,690$ connection slots that are probably filled by unreachable peers (\cref{fig-unreachable-distribution}).

To estimate the number of unreachable peers from the number of connections of unreachable peers, we need to know the number of outgoing connections of unreachable peers.
We determine the distribution of clients used by unreachable peers by calculating the distribution of user agents that are announced to our reachable peers.
Based on their distribution and the default number of outgoing connections created by each client\footnote{Bitcoin Core: 10 (8 for full relay and 2 block relay only) / 78.4\,\%, BitcoinJ: 12 / 6.5\,\%, Bread: 3 / 3.3\,\%, bcoin: 8 / 2.8\,\%},
we calculate that unreachable peers open on average 9.8 outgoing connections. 
This result leads to an estimated number of $32,800$ unreachable peers.
This estimate for the number of unreachable peers at one point in time is plausible, given that previous work has estimated the existence of 27,000 to 35,000 unreachable peers per day \cite{grundmann_announcements_2022} or 155,000 peers in each six-hour interval and a high churn \cite{wang_towards_2017}.

\section{Conclusion}

We have shown the current peer degree distribution of the Bitcoin P2P network.
As the openness of the P2P network depends on reachable peers accepting incoming connections, it is a notable result that more than a quarter of reachable peers do not accept incoming connections.
Our analysis is based on the observation of a spam wave of addresses in July and August 2021.
The same spam wave is not possible anymore in the Bitcoin P2P network because, right before the spam wave started, a change that could reduce the impact of such spam by rate-limiting \textsc{addr} propagation was implemented \cite{web-bitcoin-pr} and has been released with Bitcoin Core 22.0 \cite{web-bitcoin-version-history} in September 2021.
However, the insights that our observations of the spam wave gave into the Bitcoin P2P network can be helpful for future development of Bitcoin and other P2P networks.

\bibliographystyle{splncs04}
\bibliography{library-clean,websites}
\end{document}